\documentclass[a4paper]{article}

\usepackage[T1]{fontenc}
\usepackage[utf8x]{inputenc}
\usepackage[english]{babel}

\usepackage{float}

\usepackage{siunitx}

\usepackage[colorlinks=true, allcolors=blue]{hyperref}

\newcommand{\email}[1]{\href{mailto:#1}{\tt{\nolinkurl{#1}}}}
\newcommand{\orcid}[1]{ORCID: \href{https://orcid.org/#1}{\tt{\nolinkurl{#1}}}}

\usepackage{fancyhdr}
\usepackage{authblk}
\usepackage[a4paper,top=3cm,bottom=2cm,left=3cm,right=3cm,marginparwidth=1.75cm]{geometry}

\usepackage{amsmath}
\usepackage{graphicx}
\usepackage{booktabs}

\usepackage[colorinlistoftodos]{todonotes}

\author{C. Javaherian}
\author{C. Ferrie}
\affil{Centre for Quantum Software and Information, University of Technology Sydney, Australia} 

\usepackage[english]{babel}
\usepackage[utf8x]{inputenc}
\usepackage[T1]{fontenc}
\usepackage[a4paper,top=3cm,bottom=2cm,left=3cm,right=3cm,marginparwidth=1.75cm]{geometry}
\usepackage{amsmath}
\usepackage{graphicx}
\usepackage[colorinlistoftodos]{todonotes}
\usepackage[colorlinks=true, allcolors=blue]{hyperref}
\usepackage{braket}
\usepackage{appendix}
\usepackage{physics}
\usepackage{amssymb}
\usepackage{bbm}
\usepackage{mathtools}

\usepackage{chngcntr}
\begin{document}
\title{Energy transport and optimal design of noisy Platonic quantum networks}

\maketitle
\section*{Abstract}
Optimal transport is one of the primary goals for designing efficient quantum networks. In this work, the maximum transport is investigated for three-dimensional quantum networks with Platonic geometries affected by dephasing and dissipative Markovian noise. The network and the environmental characteristics corresponding the optimal design are obtained and investigated for five Platonic networks with 4, 6, 8, 12, and 20 number of sites that one of the sites is connected to a sink site through a dissipative process. Such optimal designs could have various applications like switching and multiplexing in quantum circuits.

\section*{Introduction}
Transport is an essential phenomena in atomic-scale devices and networks. The structure that hosts the energy or information carriers could be a continuous medium like metallic nanorods and waveguides \cite{Javaherian_2009, Javaherian_2009_2} or site-based structure like metallic nanoparticle arrays \cite{LI2018213, maier_2003}, quantum dot arrays \cite{Braakman2013, Wang2020}, and many more.
Discrete or site-based transport --- considered in this work --- has many applications such as quantum state transport through spin chains \cite{PhysRevLett.92.187902, PhysRevA.72.012319, PhysRevA.69.052315, PhysRevA.71.052315}, quantum energy transport in chains of trapped ions, environment-assisted transport in networks of sites \cite{PhysRevA.90.042313, PhysRevA.83.013811}, and switching with qubit arrays \cite{PhysRevLett.105.167001}. While much is known for ideal networks, the effect of noise on the desired properties are less understood. We study the case of noisy quantum networks here.

A specific type of quantum network that has been proven to have exact theoretical solutions for site-based energy excitation transfer are called Fully Connected Networks (FCNs) \cite{J.Chem.Phys.2009}. FCNs are defined by the property that all sites are equally connected to each other and the last site is dissipatively connected to a sink site. In \cite{Javaherian2015} we studied some three dimensional Platonic configurations with distance dependent couplings, and proved that they have some similar properties as those of an $N$-site FCN, where in the corresponding Platonic network $N-1$ would be the number of nearest neighbours of each site. For example, it was shown that the sink population --- the energy excitation accumulated in the sink site --- of the "noiseless" Platonic quantum networks and FCNs are the same at the steady state (or infinite time). These identities are convenient and powerful tools for the study quantum networks, and we find similar relations in the more relevant case of \emph{noisy} quantum networks.

In this work we prove that in different noisy Platonic quantum networks the analytical solution of the steady sink population is the same as that of the equivalent FCN. In addition, we provide some relations among the couplings and noise rates corresponding to the maximum transport. These relations will be useful to optimally design such quantum networks for various applications such as transport through three dimensional Platonic networks, or switching and multiplexing in quantum circuits. The latter could be achieved by changing the environmental noise rates (electrical or magnetic) around the last site connected to the sink site, in a quantum dot or photonic qubits implementations, so that the incoming excitation would be transferred to the planned sink site.

Before presenting our results, we comment on two advantages of using Platonic quantum networks for such purposes. First, three dimensional networks in general
%, apart from the difficulty in being constructed by existing methods, 
are more compact in comparison with two and one dimensional networks,
and the nanoscale 3D
printing methods \cite{doi:10.1021/acs.nanolett.1c02847, doi:10.1038/s42005-021-00532-4} could help to overcome the 3D construction difficulties. The three dimensional structures might also be compatible with the physics or other constraints of some architectures. The other advantage of Platonic quantum networks is that they are proven to be a FCN network with reduced number of sites that has exact transport solutions, and thus ideal benchmarks for new techniques and demonstrations.

%\section*{Noisy Platonic networks connected to a sink- energy transport and optimal design}
\section*{Results}
\begin{figure}[ht]
\center
\includegraphics[width=0.5\linewidth]{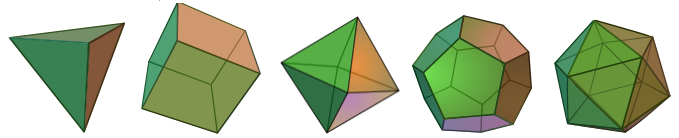}
\caption{The schematic of the five Platonic geometries. Left to right: Tetrahedron (4 vertices), Cube (8 vertices), Octahedron (6 vertices), Icosahedron (12 vertices), and Dodecahedron (20 vertices)}.
\label{Fig0}
\end{figure}

A Platonic network in this work is defined as a group of interactive identical two-level systems, located on the vertices of one of the five Platonic geometries, depicted in Fig.\ref{Fig0}, and one additional site (sink site) is irreversibly connected to one of the main sites with rate $\Gamma$. We assume a homogeneous environment so that all sites (main qubits) are affected by equal dissipation and dephasing Markovian local noises with rates $\Gamma_{\rm diss}$ and $\gamma$, respectively.

%section an analytical expression will be calculated for the final amount of quantum energy transferred through a Platonic network and accumulated in a target site in the vicinity of a particular site of the network. Then an analysis will be provided for the complete transfer of the quantum energy that could be used for optimal design of various Platonic networks with different architectures. 

%We consider the Platonic networks as depicted in Fig. \ref{Fig.1}: Tetrahedron with four  sites, Cube with eight sites, and icosahedron with twelve sites. We assume that all sites are identical two-level systems located on the vertices of the platonic configurations and are initialized in the ground state except the first site that is in the excited state. 

%The last indexed site which is assumed to be a nearest neighbor of the first site is irreversibly connected to a target sink with rate $\Gamma$. Moreover, we assume that all sites excluding the target sink are equally affected by a Markovian dissipation and dephasing noises with rates $\Gamma_{\rm diss}$ and $\gamma$, respectively.\\

%We numerically found the populations of the target sites of the Platonic networks in the lossless case \cite{Javaherian2015}. Fig.2 of \cite{Javaherian2015} indicates that at equilibrium (steady state), the value of target populations are identical for nearest and non-nearest neighbor interacting Platonic networks, i.e. networks in which all sites are coupled to each other (non-nearest neighbor interacting network), any site is only coupled to its closest sites and is negligibly coupled to other sites (non-nearest neighbor network).

%If Markovian local noises present, one might expect to have mitigated couplings between pairs of sites. So, in this work, 

%we assume the nearest neighbor approximation to calculate the target population of \textit{noisy} Platonic networks in the steady state. \\

The Hamiltonian (in units $\hbar=1$) of the Platonic network would be 
\begin{equation}
H= \sum_{i=1}^{N} J_{ij}\sigma_i^{\dagger} \sigma_j + c.c.,    
\end{equation}
where $N$ is the total number of main sites, $J_{ij}$ is the distance-dependent coupling between qubits $i,j$,  $\sigma_i^{\dagger}$($\sigma_i$) is the creation (annihilation) operator, and $c.c.$ denotes the complex conjugation of the fist part.
%$N_c-1$ is the coordinate number of a Platonic network that is equal to the number of nearest neighbors of a site, $f(N,N_c)$ is a set with a maximum number of indices of sites that do not have any common nearest neighbors, $f_{\rm N_c}(i)$ is the set of indices of $N_c$ nearest neighbors of site $i$, $J$ is the identical coupling constant between nearest neighboring sites,
The dynamics of the total density matrix of this system is found by a Lindbladian master equation \cite{Breuer2007} as follows:
\begin{align}\label{equationset1}
\qquad \dot{\hat{\rho}}&=-i[\hat{H}_{\rm sys},\hat{\rho}]+{\cal{L}}_{\rm target}(\hat{\rho})+{\cal{L}}_{deph}(\hat{\rho})+{\cal{L}}_{\rm diss}(\hat{\rho});
 \nonumber \\
\qquad {\cal{L}}_{\rm target}(\hat{\rho}) &= \Gamma(2\hat{\sigma}_{\rm target}^{\dagger}\hat{\sigma_{N}}^{}\hat{\rho}\hat{\sigma_{N}}^{\dagger}\hat{\sigma}_{\rm target}^{} -\left\{ \hat{\sigma_{N}}^{\dagger}\hat{\sigma}_{\rm target}^{}\hat{\sigma}_{\rm target}^{\dagger}\hat{\sigma_{N}}^{},\hat{\rho}\right\}),
 \nonumber \\
\qquad {\cal{L}}_{deph}(\hat{\rho})&=\gamma \sum_{i=1}^N(2\hat{\sigma_{i}}^{\dagger}\hat{\sigma_{i}}\hat{\rho}\hat{\sigma_{i}}^{\dagger}\hat{\sigma_{i}}-\left\{ \hat{\sigma_{i}}^{\dagger}\hat{\sigma_{i}},\hat{\rho}\right\}),
 \nonumber\\
\qquad {\cal{L}}_{\rm diss}(\hat{\rho}) &= \Gamma_{\rm diss}(2\hat{\sigma}_{\rm target}^{\dagger}\hat{\sigma_{N}}^{}\hat{\rho}\hat{\sigma_{N}}^{\dagger}\hat{\sigma}_{\rm target}^{} -\left\{ \hat{\sigma_{N}}^{\dagger}\hat{\sigma}_{\rm target}^{}\hat{\sigma}_{\rm target}^{\dagger}\hat{\sigma_{N}}^{},\hat{\rho}\right\}),
\end{align}
where we assumed  
$\hat{\rho}=\hat{\rho}_{\rm qubits} \oplus \hat{\rho}_{\rm target} \oplus \hat{\rho}_{\rm discharge}$, is the direct sum of the $N\times N$ density matrix of the Platonic network ($\hat{\rho}_{\rm  qubits}$), the $1\times 1$ population matrix of the sink site ($\hat{\rho}_{\rm target}$), and the $1\times 1$ matrix representing the total population discharged to the environment by dissipation noise ($\hat{\rho}_{\rm discharged}$). $\gamma$, and $\Gamma_{\rm diss}$ are the dephasing and dissipation noise rates to the environment from the network sites to the surrounding environment. $\Gamma$ is the rate of irreversible energy transfer from the last site to the target site
and $\hat \sigma_{i}^{\dagger}(\hat \sigma_{i})$ is the creation (annihilation) operator of site $i$.
%(Eq.1 of \cite{Javaherian2015}), 
%The time evolution of the total density matrix of the system including N sites of the Platonic network and a sink site as the $(N+1)^{\rm st}$ site, in the Markovian regime is a Lindbladian master equation i.e. Eq. (4) of \cite{Javaherian2015}. 
We aim to provide an analytical expression for the population of target sink i.e. $\rho_{\rm (N+1),(N+1)}(t)=\rho_{\rm target}(t)$ in the equilibrium state i.e. $t\rightarrow \infty$.

\begin{figure}[ht]
\center
\includegraphics[width=0.7\linewidth]{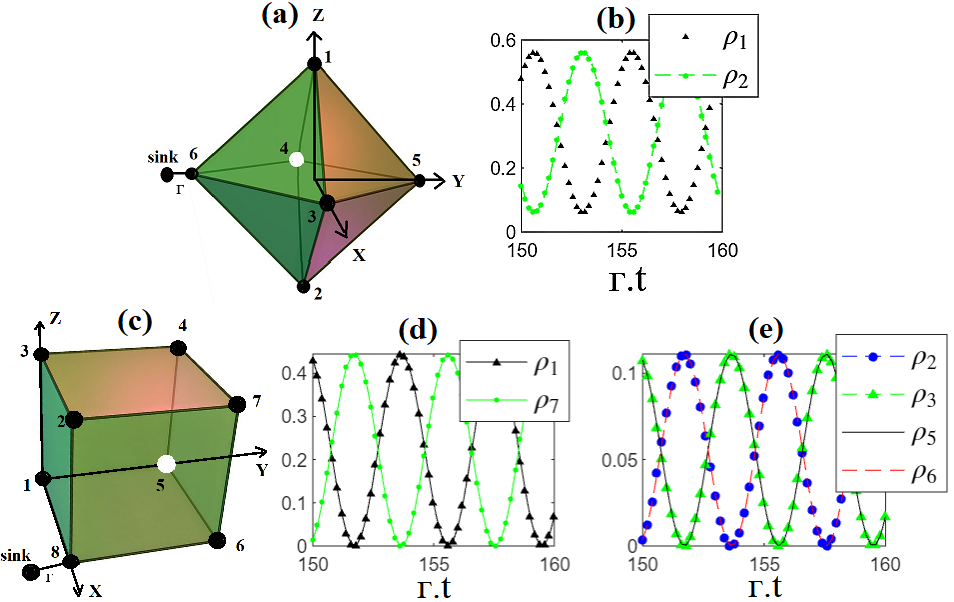}
\caption{This graph shows the simulated population dynamics of all sites ($\rho_{ii}\equiv \rho_{i},\:i=1...N=6,8$) of tetrahedron and cube noiseless networks with zero environmental noises ($\gamma=\Gamma_{diss}=0$), where the sink site ($i=N+1$) is dissipatively connected to the $N^{th}$ site with rate $\Gamma$. Parts (a) and (b) show the schematics of the tetrahedron and cube networks with 6 and 8 sites on the vertices. The sites coordinates for tetrahedron are $(0,0,\pm\frac{1}{\sqrt{2}}),(0,\pm\frac{1}{\sqrt{2}},0),(\pm\frac{1}{\sqrt{2}},0,0)$, and for the cube (graph (c)), site 1 is located at the origin of coordinates and its symmetrically positioned site (7) is located at $(1,1,1)$. In both networks, at $t=0$, site No. 1 was charged by one excitation. Part (b) presents the oscillating populations of sites 1 and 2 in time that are positioned symmetrically with respect to the centre of the tetrahedron. Note that since the rate $\Gamma$ has dimension $1/T$, the quantity $\Gamma.t$ is dimensionless. The population of the spherically symmetric sites (3,4) at equilibrium i.e. $\Gamma.t \gtrsim 20$ is constant ($0.0625$). The symmetrically positioned sites 5 and 6 are discharged at equilibrium while the sink site 7 is saturated to the population of $0.25$. The graphs (d) and (e) show that the spherically symmetric sites (positioned symmetrically with respect to the centre of the cube), i.e. (1,7), (2,5) and (3,6) are oscillating inversely at equilibrium i.e. $\Gamma.t \gtrsim 20$. At equilibrium, the populations of the spherically symmetric sites 4 and 8 are zero since the population of the sink site (9) is saturated to $0.33$.}
\label{N6-8}
\end{figure}

\begin{figure}[ht]
\center
\includegraphics[width=0.9\linewidth]{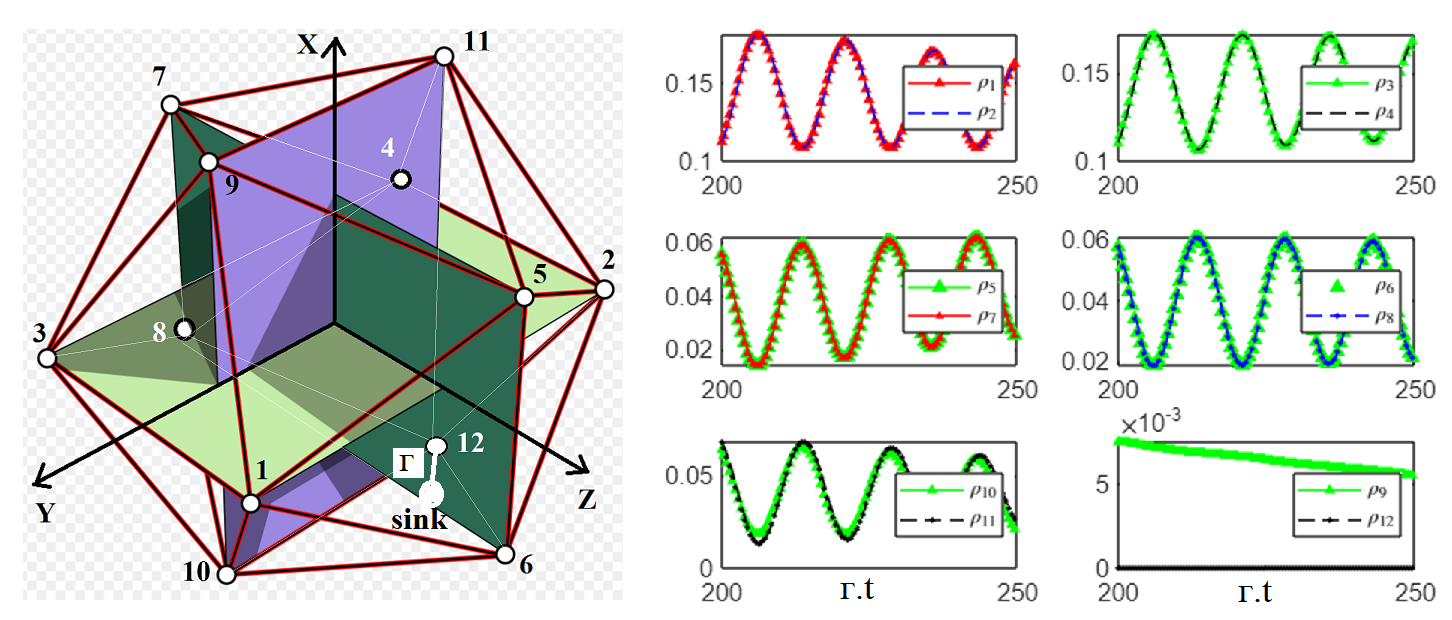}
\caption{This graph shows the simulated population dynamics of all sites ($\rho_{ii}\equiv \rho_{i},\:i=1...N=12$) of a regular icosahedron noiseless network with zero environmental noises ($\gamma=\Gamma_{diss}=0$), where the sink site ($i=N+1=13$) is dissipatively connected to only one site (12) with rate $\Gamma$. The left graph shows the schematic of the icosahedron network where sites are located on the vertices with coordinates $(0,\pm \phi, \pm 1),(\pm 1, 0, \pm \phi),(\pm \phi, \pm 1, 0)$ where $\phi = (1+\sqrt(5))/2$. At $t=0$ four sites (1,2,3,4) are equally charged by $1/4$ amount of excitation. The right graphs show that due to the various symmetries in Platonic geometries, some sites are oscillating similarly (sites "1,2", "3,4", "5,7", "6,8", and "10,11"), at the equilibrium i.e. $\Gamma.t \gtrsim 20$. In addition, the populations of the spherical-symmetrically positioned sites of 12 and 9, are zero, since the population of the sink site (13) is saturated to $0.2$ at equilibrium.}
\label{N12}
\end{figure}

\begin{figure}[ht]
\center
\includegraphics[width=0.9\linewidth]{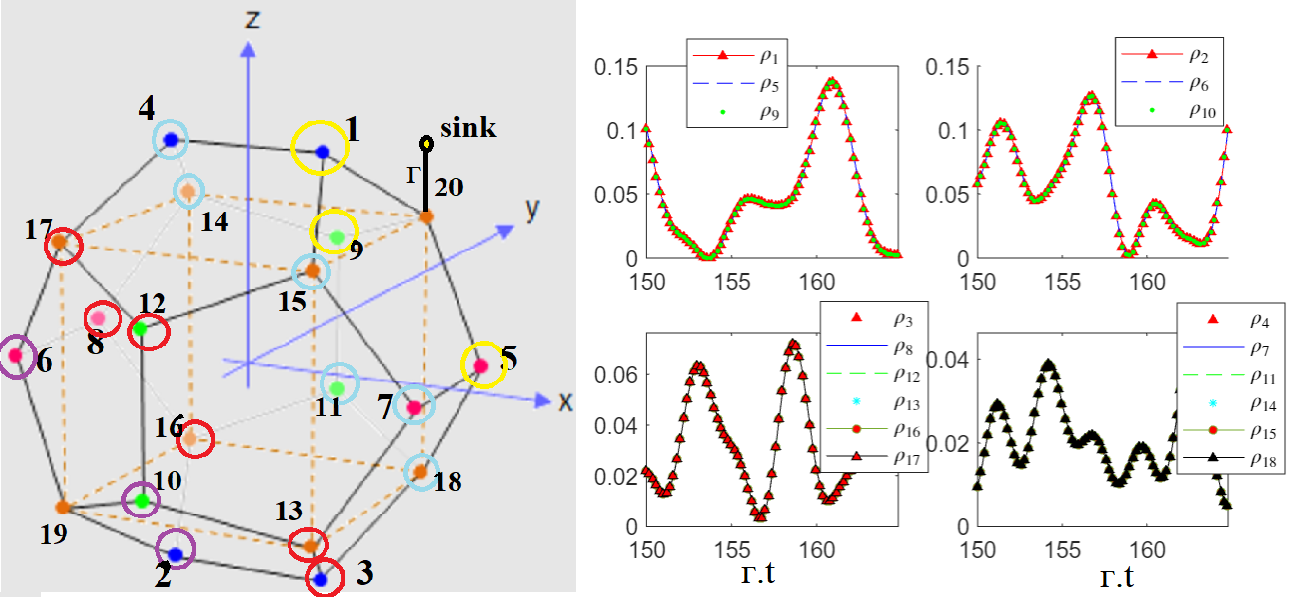}
\caption{This graph shows the simulated population dynamics of all sites ($\rho_{ii}\equiv \rho_{i},i=1...N=20$) of a regular dodecahedron noiseless network with zero environmental noises ($\gamma=\Gamma_{diss}=0$), where the sink site ($i=N+1=21$) is dissipatively connected to only one site (20) with rate $\Gamma$. The left graph shows the schematic of the icosahedron network where sites are located on the vertices with coordinates $(\pm 1, \pm 1, \pm 1), (0,\pm \phi, \pm \frac{1}{\phi}),(\pm \phi, 0, \pm \frac{1}{\phi}),(\pm \phi, \pm \frac{1}{\phi}, 0),$ where $\phi = (1+\sqrt(5))/2$. At $t=0$ three sites (1,5,9) are equally charged by $1/4$ amount of excitation. The right graphs show that due to the various symmetries in Platonic geometries, some sites are oscillating similarly (sites "1,2", "3,4", "5,6,7,8", and "10,11"), at the equilibrium i.e. $\Gamma.t \gtrsim 20$. In addition, the populations of the spherically symmetric sites 19 and 20 are zero since the population of the sink site (21) is saturated to $0.33$. There are four groups of sites on the right-side schematics that are marked with four different colors (violet, red, light blue, and yellow), that each group of sites posses similar populations. Adding two sites to red and yellow groups (sites No. 19 and 20, respectively), we create the four symbolic sites of the equivalent FCN of dodecahedron, as introduced in Eq.\eqref{N20FCNsites}}
\label{N20}
\end{figure}

In order to simplify the analytical calculation, we study the dynamical characteristics of Platonic networks by numerically solving $\rho(t)$ from Eq.\eqref{equationset1}. For the simulation, the ion qubits implementation of the Platonic networks is assumed in which the coupling rate, or the interaction energy between two dipoles of ion qubits $i$ and $j$ would be inversely proportional to the cube of their interconnecting distance ($J_{ij}=v/r^3_{ij},\:v=1$).
In Figs.\ref{N6-8},\ref{N12},\ref{N20} we plotted the sites' populations ($\rho_{ii}(t)$) of four Platonic networks in the noiseless case ($\gamma= \Gamma_{diss}=0$). It can be seen that the populations of some sites vary inversely in some cases, while other sites would have same population dynamics.   
The characteristics shown in these figures (noiseless cases) would be the same as that of noisy cases unless the fact that in the presents of dephasing and/or dissipation noises, the oscillating patterns of populations would be evanescent and the sink site would be fully populated at the equilibrium ($t \rightarrow \infty$).
%Figs. \ref{N12} and \ref{N20} show the dynamcis of two other cases N=12 and 20. It can be seen that all pairs of sites in these two cases are varying similarly. 
In \cite{Javaherian2015} we found that the target population of noiseless Platonic networks at steady state are independent of their size and is only related to the number of neighboring sites $(N_c-1)$ as of $\rho_{sink}=1/(N_c-1)$, i.e. $\rho_{sink}=0.25,0.33,0.2,0.33$ for $N=6,8,12,20$, respectively. 
Later on, we conclude that the exact solution of the dynamics of Platonic networks are the same as that of FCNs, in which all sites are equidistant from each other \cite{J.Chem.Phys.2009}. Since the tetrahedron Platonic network (N=4) is a FCN by definition, we ignored its simulation since we do not need anymore to prove that its analytical solution of target population is as that of an FCN.

Figs.\ref{N6-8} show the dynamics of two Platonic networks with N=6 and 8. It can be seen that some pair of sites are oscillating inversely, while the populations of other pairs of sites vary equally. 
To simplify the analytical solution, we will assume that the average of sites' populations (and the coherences) of these pairs could be assumed for one of the sites, and we would only solve the dynamics of a Platonic network for $N_c$ sites.
Figs.\ref{N12} and \ref{N20} show the dynamics of Platonic networks with N=12 and N=20 sites, respectively. The graphs show that in these types of networks, for the chosen initial charges of each case, the populations of different groups of sites vary similarly. So likewise the other Platonic geometries, we simplify the analytical dynamics by assuming only a specific number of sites i.e. $N_c$ sites.
So if $\rho(t)$ would be the density matrix of a platonic network of $N$ sites, with rank $(N+2)$ and elements $\rho_{ij}(t)$, we define the following symbolic density matrix $\tilde{\rho}(t)$ of an equivalent network of $N_c$ sites, with rank $(N_c+2)$ and symbolic elements of $\tilde{\rho}_{ij}(t)$ defined as follows. Note that in the legends of Figs. \ref{N6-8},\ref{N12},\ref{N20} and the two following formulas, we consider the notation $\rho_{ii}(t) \equiv \rho_{i}(t)$ for simplicity.

\begin{align}\label{assumption}
\begin{split}
N=6:&\\
&\tilde{\rho}_{1}(t)=  \rho_{1}(t)+\rho_{2}(t),\\
&\tilde{\rho}_{i}(t)=  \rho_{i}(t),\: i \ne 1,2\\
N=8,12:&\\
&\tilde{\rho}_{i}(t) =  
\rho_{i}(t) + \rho_{\tilde{i}}(t),\\
\end{split}
\end{align}
where $\tilde{i}$ is the index of the spherical-symmetrically positioned site with respect to site $i$. So the number of symbolic sites of $N=6$ network is $N_c=5$, and that of $N=8$ and $12$ are $N_c=4$ and $N_c=6$, respectively. 
For $N=20$, according to Fig.\ref{N20} we choose:
\begin{align}\label{N20FCNsites}
\begin{split}
N=20:&\\
&\tilde{\rho}_{1}(t)=  \rho_{1}(t)+\rho_{5}(t)+\rho_{9}(t)+\rho_{20}(t),\\
&\tilde{\rho}_{2}(t)=  \rho_{2}(t)+\rho_{6}(t)+\rho_{10}(t)+\rho_{19}(t),\\
&\tilde{\rho}_{3}(t)=  \rho_{3}(t)+\rho_{8}(t)+\rho_{12}(t)+\rho_{13}(t)+\rho_{16}(t)+\rho_{17}(t),\\
&\tilde{\rho}_{4}(t)=  \rho_{4}(t)+\rho_{7}(t)+\rho_{11}(t)+\rho_{14}(t)+\rho_{15}(t)+\rho_{18}(t),\\
\end{split}
\end{align}
So the number of symbolic sites of $N=20$ network is $N_c=4$,
The above two formulas show that we assume $N_c=5,4,6,4$ symbolic sites, for five Platonic networks with $N=6,8,12,20$ sites. 
Note that the distance between the chosen group of sites, assumed as one symbolic site, with the other neighboring group of sites are all equal. 
%the symbolic (or equivalent) network could be considered as of a FCN for the cases $N=6,8,12$. 
%For $N=20$ we will show that the steady state target population of the assumed symbolic noisy network with $N_c=4$ sites is as that of an FCN network. 
Now the coherences between the sites of the equivalent reduced networks would be defined according to the symbolic indices as 
\begin{equation}\label{coherences}
\tilde{\rho}_{ij}(t)=\sum_{p=1}^n\sum_{q=1}^m\rho_{pq}(t),
\end{equation}

where $n,m$ are the number of sites that define the symbolic sites $i,j$, respectively, that is $\tilde{\rho}_{ii}(t) \equiv \tilde{\rho}_{i}(t)=\sum_{p=1}^n\rho_{p}(t)$ and $\tilde{\rho}_{jj}(t)\equiv \tilde{\rho}_{j}(t)=\sum_{q=1}^n\rho_{q}(t)$. As an example, in a dodecahedron network:
\begin{equation}\label{coherences-example}
\tilde{\rho}_{12}(t)=\sum_{\substack{p=\{ 1,5,9,20\},\\
q=\{2,6,10,19\}}}\rho_{pq}(t).
\end{equation}

In future, for simplicity of notation, we substitute $\tilde{\rho}(t)\rightarrow \rho(t)$.
Using the assumption of reduced networks, Eq.\eqref{equationset1} would be written as follows: 
\begin{equation} 
\label{eq1}
\begin{split}
\dot{\rho}_{ii} &= -2\Gamma_{\rm diss}\rho_{ii}+iJ (R_i-\bar{R}_i) ;\:\: i\ne N,\\
\dot{\rho}_{ij} &= -2(\Gamma_{\rm diss}+\gamma)\rho_{ij}+iJ (R_i-\bar{R}_j) ;\:\: (i,j)\ne N,\\
\dot{\rho}_{iN} &= -(2\Gamma_{\rm diss}+2\gamma+\Gamma)\rho_{iN}+iJ (R_i-\bar{R}_N),\\
\dot{\rho}_{\rm NN} &= -(2\Gamma_{\rm diss}+2\gamma+\Gamma)\rho_{\rm NN}+iJ (R_N-\bar{R}_N),\\
\dot{\rho}_{00} &= 2\Gamma_{\rm diss}Tr_{N_c}(\rho),\\
\dot{\rho}_{\rm target} & = 2\Gamma \rho_{\rm NN},
\end{split}
\end{equation}
where $\rho_{00}$ correspond to a virtual site that stores a fraction of initial excitation within the environment through dissipative noise with rate $\Gamma_{diss}$,  and the population of the last site $\rho_{NN}$, is being discharged dissipatively by rate $\Gamma$ to the target site with corresponding matrix element $\rho_{(N+1),(N+1)}$. We assume the following collective variables:
\begin{equation}
R_i(t)=\sum_{j=f_{\rm N_c}(i)} \rho_{ij}(t),\:\:\Lambda_{i}=\sum_{j=f_{\rm N_c}(i)} R_j,    
\end{equation}
where $f_{\rm N_c}(i)$ is the set of $N_c$ indices 
of the nearest neighbors of site $i$ plus itself.
%(which for our chosen assumptions and Platonic networks sites $1$ and $N$ are included in $f_{\rm N_c}(i)$
%, and $\bar{R}_i(t)$  denotes its complex conjugate. We define by defining another variable
%\begin{equation}
%\Lambda_{i}=\sum_{j=f_{\rm N_c}(i)} R_j,    
%\end{equation}
The equations of motion of the collective variables would be as following: 
\begin{equation} 
\label{eq2}
\begin{split}
\dot{R_{i}} &= -iJ\Lambda_i + i N_c J R_i - 2 (\Gamma_{\rm diss}+\gamma)R_i - \Gamma \rho_{iN} +2\gamma \rho_{ii,}\\
\dot{R_{N}} &= -iJ\Lambda_N + i N_c J R_N - (2\Gamma_{\rm diss}+2\gamma+\Gamma)R_N + (2\gamma-\Gamma) \rho_{\rm NN},\\
\dot{\Lambda_{i}} &= -(2\Gamma_{diss}+\gamma)\Lambda_i + \Gamma(R_N+\bar{R}_N) +2\gamma Tr_{\rm N_c(i)}(\rho),\\
\end{split}
\end{equation}
where
\begin{equation}
Tr_{\rm N_c(i)}(\rho)=\sum_{i=f_{\rm N_c}(i)} \rho_{ii} = Tr(\rho).    
\end{equation}
%As explained in Fig.\ref{Fig.1} for three Platonic networks with $N=4,8,12$ one obtains:
%\begin{equation}
%Tr_{\rm N_c}(\rho)=1/\lfloor N/N_c \rfloor Tr(\rho), %\end{equation}
%where index $i$ is eliminated. This is illustrated in Fig.\ref{Fig.1}, and assumes that due to the geometric symmetry and  in the equilibrium state ($t \rightarrow \infty$), any site in such networks possess the same average of excitation population as that of its mirror state. In other words, the sum of average populations are identical, for example in a Cube network (Fig.\ref{Fig.1}(b)), i.e.
%\begin{align}
%\begin{split}
%Tr_{N_c(1)}(\rho) &=\rho_{11}+\rho_{33}+\rho_{55}+\rho_{88} \\ 
%&\equiv 0.5[(\rho_{11}+\rho_{\tilde{1}\tilde{1}})+(\rho_{33}+\rho_{\tilde{3}\tilde{3}})+(\rho_{55}+\rho_{\tilde{5}\tilde{5}})+(\rho_{88}+\rho_{\tilde{8}\tilde{8}})] \\
%&\equiv \rho_{11}+\rho_{22}+\rho_{66}+\rho_{88}=Tr_{N_c(8)}(\rho).
%\end{split}
%\end{align}
Now from the rule of conservation of population we have:
\begin{equation} 
\label{eq3}
\begin{split}
1 =  Tr(\rho) +\rho_{00} + \rho_{\rm target}.
\end{split}
\end{equation}

%\begin{equation} 
%\label{eq3}
%\begin{split}
%1 =  \lfloor N/N_c \rfloor Tr_{\rm N_c}(\rho) +\rho_{00} + \rho_{\rm target}.
%\end{split}
%\end{equation}
It indicates that the initial population would oscillate among all network sites, and partially accumulated in the surrounding environment and the target site, through the dissipation noise rates of $\Gamma_{\rm diss},\gamma$ and $\Gamma$, respectively.\\

Considering $R_N = x+iy$, Eq.\eqref{eq2} line 2 yields two first order differential equations for $x$ and $y$ which besides Eqs.\eqref{eq1} lines 4-6, Eq.\eqref{eq2} line 3 and Eq.\eqref{eq3}, form a close set of differential equations for variables $x, y, \Lambda_N, \rho_{\rm NN}, \rho_{00},\rho_{\rm target}$ as follows:
\begin{equation} 
\label{eq4'}
\begin{split}
\dot{\Lambda}_N &= -2(\Gamma_{diss}+\gamma)\Lambda_N -2\Gamma x + 2\gamma (1-\rho_{00}-\rho_{\rm target}),\\
\dot{x} &= -(2\Gamma_{diss}+2\gamma+\Gamma)x + (2\gamma-\Gamma)\rho_{\rm NN} -JN_cy,\\
\dot{y} &= -(2\Gamma_{diss}+2\gamma+\Gamma)y + JN_cx -J \Lambda_N,\\
\dot{\rho}_{\rm NN} &= -2(\Gamma_{diss}+\Gamma)\rho_{\rm NN}-2Jy,\\
\dot{\rho}_{00} &= 2\Gamma_{diss} (1-\rho_{00}-\rho_{\rm target}),\\
\dot{\rho}_{\rm target} &= 2\Gamma\rho_{\rm NN}.
\end{split}
\end{equation}

%\begin{equation} 
%\label{eq4'}
%\begin{split}
%\dot{\Lambda} &= -2(\Gamma+\gamma)\Lambda -2\Gamma x + 2\gamma C_p (1-\rho_{00}-\rho_{\rm target}),\\
%\dot{x} &= -(2\Gamma+2\gamma+\Gamma)x + (2\gamma-\Gamma)\rho_{\rm NN} -JN_cy,\\
%\dot{y} &= -(2\Gamma+2\gamma+\Gamma)y + JN_cx -J \Lambda,\\
%\dot{\rho}_{\rm NN} &= -2(\Gamma+\Gamma_N)\rho_{\rm NN}-2Jy,\\
%\dot{\rho}_{00} &= 2\Gamma C_p(1-\rho_{00}-\rho_{\rm target}),\\
%\dot{\rho}_{\rm target} &= 2\Gamma_N\rho_{\rm NN}.
%\end{split}
%\end{equation}

According to Eq.\eqref{assumption} and Figs.\ref{N6-8},\ref{N12},\ref{N20}, the initial conditions for different Platonic networks are assumed as:
\begin{align}\label{initial-conditions}
\begin{split}
N=6,8:&\\
&\rho_{1}(0)\equiv \rho_{11}(0)=1,\:\rho_{ij}(0)=0,i,j \ne 1\\
N=12:&\\
&\rho_{1}(0)=\rho_{2}(0)=\rho_{3}(0)=\rho_{4}(0)=1/4,
\:\rho_{ij}(0)=0,\:i \ne j \ne 1,2,3,4\\
N=20:&\\
&\rho_{1}(0)=\rho_{5}(0)=\rho_{9}(0)=1/3,
\:\rho_{ij}(0)=0,\:i \ne j \ne 1,5,9
\end{split}
\end{align}
%$\rho_{ij}(t=0)=\delta_{i1}\delta_{j1}$.
The above initial conditions yield:
\begin{align}
\begin{split}
R_N(0)  &= \sum_{i\in f_{\rm N_c}(N)} \rho_{Ni}=0, \:x(0)=y(0)=0,\\
\Lambda_N(0) &=\sum_{i\in f_{\rm N_c}(N)} R_i(0)=R_N(0)+R_{i\ne\{1-4,N\}\:\rm or,i\ne \{1,5,9,N\}}(0)=1.     
\end{split}
\end{align}
Now by applying the Laplace transform to Eqs.\eqref{eq4'} i.e. $t\rightarrow 1/s$, $\dot{\alpha}(t) \rightarrow \Big(\tilde{\alpha}.s -\alpha(0)\Big)$, we obtain: 
\begin{equation} 
\label{eq5'}
\begin{split}
(s+2\Gamma_{diss}+2\gamma)\tilde{\Lambda}_N + 2\Gamma \tilde{x} + 2\gamma \rho_{\rm target} +  2\gamma \rho_{00} - 2\gamma /s -1 =0,\\
(s+2\Gamma_{diss}+2\gamma+\Gamma)\tilde{x} + (\Gamma-2\gamma)\tilde{\rho}_{\rm NN} + JN_c\tilde{y} = 0,\\ 
(s+2\Gamma_{diss}+2\gamma+\Gamma)\tilde{y} + J\tilde{\Lambda}_N -JN_c\tilde{x} = 0,\\ 
(s+2\Gamma_{diss}+2\Gamma)\tilde{\rho}_{\rm NN} +2J\tilde{y} = 0,\\
(s+2\Gamma_{diss})\tilde{\rho}_{00} + 2\Gamma_{diss} \tilde{\rho}_{\rm target} -2\Gamma_{diss}/s = 0,\\
s\tilde{\rho}_{\rm target} - 2\Gamma\tilde{\rho}_{\rm NN} = 0.
\end{split}
\end{equation}
 
%The factor $C_p$ in Eq. \eqref{eq5'} is identified by the type of Platonic network, and is an additional term to the equivalent Eq.(A9) of \cite{J.Chem.Phys.2009}, that studied fully connected networks.\\
Solving the complete set of Eqs.\eqref{eq5'}, the target sink population will be found for 
%three 
Platonic networks 
%($N=4,8,12$) 
in the presence of homogeneous local noises as following: 

\begin{equation} \label{eq4}
\begin{split}
\bar{\rho}_{\rm target}=  4 \Gamma J ^ 2 \frac{ (\Gamma_B + s) (s + \Gamma_A)}{s \Delta(s)}\\
\Delta(s)=   8 \Gamma J ^ 2 \gamma (\Gamma_B + s) +  (s + 2 \Gamma_{\rm diss})\\
                   ( (\Gamma_A + s) (\Gamma_C + s) (\Gamma_B + s) ^ 2\\ 
                   - 4 \Gamma J ^ 2 (\Gamma - 2 \gamma) \\
                  - 2 J ^ 2 N_c (\Gamma_A + s) (\Gamma - 2 \gamma)\\ 
                  + 2 \Gamma J ^ 2 N_c (\Gamma_C + s) \\
                  + J ^ 2 N_c ^ 2 (\Gamma_A + s) (\Gamma_C + s)) .
\end{split}
\end{equation}

where 
\begin{align}
\begin{split}
\Gamma_A &=2\gamma+2\Gamma_{\rm diss}, \\ 
\Gamma_B &=2\gamma+2\Gamma_{\rm diss}+\Gamma,\\
\Gamma_C &=2\Gamma_{\rm diss}+2\Gamma
\end{split}
\end{align}
This expression is equivalent to that of an FCN network, i.e. Eqs.(A25, A26) of \cite{J.Chem.Phys.2009}, where the Platonic coordinate number ($N_c$) is equivalent to the total number of sites.
Since the root of the denominator in Eq.\eqref{eq4} does not have an analytical solution, the final target population of the considered Platonic networks are as following:

\begin{equation} 
\label{eq5}
\begin{split}
\rho_{\rm target}(t \rightarrow \infty) = \lim_{s \rightarrow 0} [s \bar{\rho}_{\rm target}] =  4 \Gamma J ^ 2 \frac{\Gamma_B \Gamma_A}{ \Delta(0)}
\end{split}
\end{equation}
It can be seen that in the noiseless environment ($\gamma, \Gamma \rightarrow 0$), the steady state target population is the same as the previous expression found in \cite{Javaherian2015} i.e. $1/(N_c -1)$, which can be here achieved by first tending the local dephasing rate to zero. If first tending the local dissipation rate to zero, the target population tends to $1$. This is due to the fact that the local dissipation noise would only discharge the excitation from each site to the environment and not to the other sites, however the dephasing noise provides new paths of energy transport within network sites, leading to discharge of all excitation to the target site through the $N^{th}$ site.

\begin{figure}[ht]
\center
\includegraphics[width=0.9\linewidth]{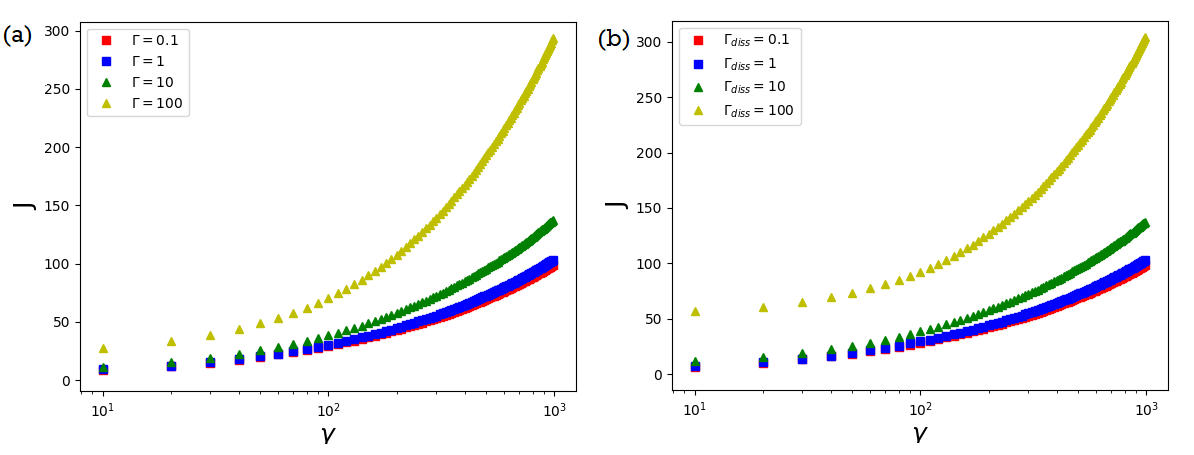}
\caption{The relation of network and environment variables in a noisy cubic Platonic network with optimal transport to the target site at steady state. Graph (a) shows the relation of coupling rate $J$ and dephasing noise $\gamma$ for different amounts of dissipation rate to the sink ($\Gamma$), while dissipation rate to the local environments are fixed at $\Gamma_{\rm diss}=10$. Graph (b) shows the same relations for different values of $\Gamma_{\rm diss}$ and fixed amount of $\Gamma=10$. It could be indicated from the graphs that the coupling strength should be increases by increasing other parameters to maintain the full transport. The values of parameters could be used for optimal design of Platonic networks.}
\label{Fig.2}
\end{figure}

\begin{figure}[ht]
\center
\includegraphics[width=0.6\linewidth]{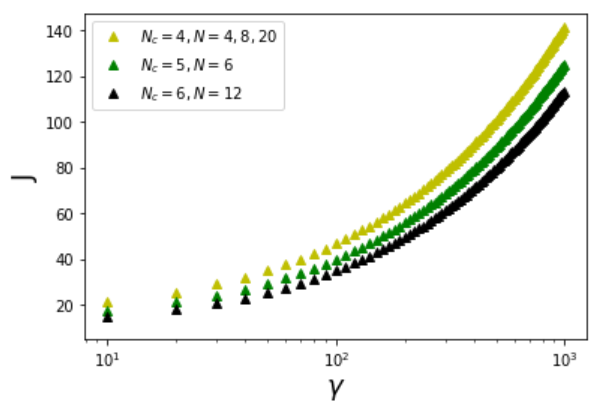}
\caption{This graph shows the relation of sites’ couplings and the environmental dephasing rate for different Platonic networks with optimal transport at steady state. It could be seen that for networks with more $Nc$ or number of nearest neighbours, the coupling rate $J$ of each pair of sites is less. This indicates that more sites with less coupling strength are equivalent to less number of sites with higher coupling rates. }
\label{Fig.3}
\end{figure}

From the numerical investigations we know that the maximum value of the target population in the presence of noises is one. To find the network-environment parameters corresponding the maximum excitation transport, we equate the target population of Eq.\eqref{eq5} to one, and find a relation among all network and environment parameters. The resulting equation can be used to find the optimal design variables. For example, the coupling rate $J$ could be found in terms of other parameters. Figs.\ref{Fig.2}(a),(b) show the relation between $J$, the coupling rate between the nearest neighbors in Platonic networks, and the Markovian dephasing rate $\gamma$, for different values of dissipation rates from each site to the environment ($\Gamma_{\rm diss}$), and also the different values of dissipation rates from the $N^{th}$ site to the sink site ($\Gamma$). The network of consideration for both parts (a) and (b) is a cubic lattice with $N=8$ main sites, and the constant parameters are $\Gamma_{\rm diss}=10$ and $\Gamma =10$, respectively. It can be seen in Fig.\ref{Fig.2} that for the fixed chosen parameters, by increasing the dephasing rate, the coupling strength should be increased, so that the disturbing effect of dephasing noise would be compensated on energy transport towards the sink. 

It can be also seen from Fig.\ref{Fig.2}(a) that for a fixed depahsing noise rate $\gamma$ and the fixed chosen dissipation noise rate $\Gamma_{\rm diss}=10$, to maintain the maximum transport, the nearest neighbors sites couplings should be increased, by increasing the dissipation rate to the target sink ($\Gamma$). In other words, in the fixed environment with the same dephasing and dissipation noise rates, the higher sites' couplings demands faster noise rates to the sink site to obtain the optimal design corresponding the maximum energy transfer. To understand this behaviour, note that the higher sites' coupling rate results in stronger and more energy bouncing among the sites, that demands higher coupling rate towards the sink. It should be also taken into account that the relation among the parameters for the maximum transport is nonlinear. 

In such Platonic noisy networks, Since the nonzero dissipation rate of $\Gamma_{\rm diss}=10$ will irreversibly transfer some energy to the environment, to reach the full energy transport, the coupling rate and the sink dissipation rate should be high or fast enough to transfer all energy before any fraction of that would be dissipated towards the surrounding environment. 
Part of this process could be supported by dephasing-noise-assisted-transport \cite{PhysRevA.90.042313, PhysRevA.83.013811}. 
This fact can be seen in Fig.\ref{Fig.2}(b). It shows the relation of network-environment parameters for the fixed amount of $\Gamma=10$. It can be seen that for a strong dissipation rate ($\Gamma_{\rm diss}=100$), the sites' coupling rate should be increased to be able to transfer the energy fast enough before getting dissipated to the environment. 

Fig.\ref{Fig.3} shows the optimal design graphs of Platonic networks with different number of sites. The constant parameters are chosen as $\Gamma_{\rm diss}=\Gamma=10$. It can be seen that for a fixed dephasing rate, by increasing $Nc$ of a Platonic network (the number of nearest neighbours of any site plus one), the coupling rate $J$ should be decreased to obtain maximum transport. This could be understood by the fact that by increasing $N_c$,  the number of interactions by a single site will be increased. So for given fixed initial excitation (one unit) to each Platonic network, we need less coupling rate between each pair of sites, to maintain maximum transport. 
In summary, Figs.\ref{Fig.2} and \ref{Fig.3} provide information to design 3D noisy Platonic networks that can fully transfer the energy from the designated sites towards a target sink. Using these optimal design graphs help us to prevent the loss of energy excitation into the surrounding Markovian environment through local dissipation and dephasing noise rates.  

\section*{Conclusion}
Energy transport is an inevitable phenomenon in many atomic-scale networks. In this work, we numerically studied the characteristics of energy dynamics in Platonic quantum networks consists of $4,6,8,12$, and $20$ qubits, located on vertices of high symmetric 3D Platonic geometries. A target site was assumed to be dissipatively connected to one of the qubits. Due to the opposite or same patterns of populations' oscillations of qubits, we made an assumption of reducing the number of qubits of each network to an effective value which was equal to the number of one group of nearest neighbor sites within each network. We found the analytical expression for the target site population in the presence of environmental Markovian dephasing and dissipation noises. In addition, we investigated the optimal design characteristics of Platonic network for maximum energy transport from the first site towards the target site. We plotted the relation between the coupling strength and the dephasing noise rate corresponding the maximum transport. The optimal designs of Platonic quantum networks could have several applications like switches or multiplexers in quantum devices. In the future, we hope to further analyse energy transport in three dimensional Platonic devices and investigate their physical implementations.
%\section*{Appendix}

\bibliographystyle{unsrt}
\bibliography{References}
\end{document}